\documentclass[journal]{IEEEtran}
\usepackage{cite}
\usepackage{graphicx}
\usepackage{subfigure}
\usepackage{amssymb,amsmath,bm}
\usepackage{url}
\usepackage{color}

\usepackage[normalem]{ulem}

\begin{document}
%
\title{On the inadequacy of the logistic map for cryptographic applications}
\author{David~Arroyo,
        Gonzalo~Alvarez,
        Veronica Fernandez
\thanks{David Arroyo, Gonzalo Alvarez and Veronica Fernandez are with Instituto de
F\'isica Aplicada, Consejo Superior de Investigaciones
Cient\'ificas, Serrano 144, 28006 Madrid, Spain. \newline e-mail: \{david.arroyo, gonzalo, veronica.fernandez\}@iec.csic.es}}

\maketitle

\markboth{Actas de la X RECSI}%
{Arroyo \MakeLowercase{\textit{et al.}}: On the inadequacy of the
logistic map for cryptographical issues.}

\begin{abstract}
This paper analyzes the use of the logistic map for cryptographic
applications. The most important characteristics of the logistic map
are shown in order to prove the inconvenience of considering this
map in the design of new chaotic cryptosystems.
\end{abstract}


\section{Introduction}
Chaotic cryptography has been an important research area during the
last two decades. The properties of chaotic systems have been used
in very different ways to build new cryptosystems. All of those
proposals can be classified into two big families, which are analog
chaos-based cryptosystems and digital chaos-based cryptosystems. The
first type of chaotic cryptosystems is based on the chaotic
synchronization technique \cite{Pecora1990}, whereas digital chaotic
cryptosystems are based on one or more chaotic maps in such a way
that the secret key is either given by the control parameters and
the initial conditions or determines those values. The scope of this
paper is related to a subclass of the last type of chaotic
cryptosystems. More specifically, the present work deals with the
underlying problems of those digital chaotic cryptosystems based on
the logistic map. The logistic map is the most widely used in the
designing of new digital chaotic cryptosystems
\cite{bap,kocarev01,jakimoski01,wong01,wong02,wong03,tang03,pareek03,huang05,pareek05,
pareek06,Wei2006,wei06b,pisarchik06,xiang06,gao07a,gao07b,wang07,ling07,mi07,yang08,xiang07,oliveira08,wang08}.
Some of these proposals have been totally or partially cryptanalyzed
as a consequence of not fully considering the dynamical
characteristics of the logistic map
\cite{Jakimoski:Analysis:PLA2001,alvarez03a,alvarez03b,alvarez04a,alvarez04b,li04,chengqingLi07,arroyo07c,physcon07,skrobek07}.
This paper analyzes these dynamical characteristics and explains the
problems of their application in the context of chaotic
cryptography.

\section{Chaotic maps and cryptography}
The major task of criptography is to ensure the secrecy of
information. To do so, encryption methods have to be created so the
message or plaintext is transformed into an encrypted message or
ciphertext. The transformation procedure depends on an external
parameter called key such that it is only possible to recover the
original message if that key is known. As it was required by
Kerckhoff \cite[p. 14]{menezes:book97}, the security of a
cryptosystem should depend only on its key. It is also necessary
that the ciphertext generated by the encryption procedure does not
contain enough information to guess either the plaintext or the
value of the key, which also forces ciphertexts associated to very
similar values of the secret key to be very different from each
other. This is the reason why chaotic maps have captured the
attention of cryptographers and many chaotic cryptosystems, i.e.,
cryptosystems based on chaotics maps have been proposed. The
dynamics of chaotic maps highly depends on an external value or set
of values and the initial state of the system, which reminds the
dependency of the cryptosystem on the secret key. Chaotic maps have
also a noise-like behavior that can be used to encrypt information
through substitution or masking processes. The next section
introduces the above mentioned dynamical properties of chaotic maps
and their special behavior.

\section{Discrete dynamical sytems}
A discrete dynamical system \cite{Wiggins1990} is defined by a
difference equation
\begin{equation}
    x \mapsto g(x),
\end{equation}
where $x, g(x) \in U \in \mathbb{R}^n$. Discrete dynamical systems
are usually studied by means of their temporal evolution. The
temporal evolution of a discrete dynamical system is given by the
sequence of points
\begin{equation}
    \left\{x_0,x_1,\ldots, x_n\right\},
\end{equation}
where $x_0, x_n\in U$ and $x_n$ is defined by
\begin{equation}
    x_k = g\left(x_{k-1}\right),
\end{equation}
for $k=1,2,\ldots,n$. This temporal evolution is usually called
orbit of the dynamical system and it is going to be noted as
$\gamma(x_0)$.

It is also possible that the dynamics of the discrete system depends
on an external parameter or set of parameters $\mu \in V \subset
\mathbb{R}^p$. In this case, the time evolution is
\begin{equation}
 x_{k+1}=g(\mu,x_k),
\end{equation}
and the orbit associated to a certain initial condition is
\begin{equation}
    \gamma(\mu, x_0)= \left\{x_0,x_1,\ldots,x_{n}\right\}.
\end{equation}
The next task is to categorize the influence of the initial
condition and the parameter or set of parameters on the time
evolution of the dynamical system. This goal does not require an
analytical determination of the orbit, since a qualitative analysis
gives enough information. The qualitative study is based on the
concept of \emph{invariant set}. Let $\Gamma$ be a subset of $U
\subset \mathbb{R}^n$. $\Gamma$ is an invariant set of
$x_{k+1}=g(\mu,x_k)$ if for any $x_0\in \Gamma$ it is satisfied that
$x_{k}\in \Gamma$ for any $k\in \mathbb{N} \cup \{0\}$. Indeed, the
determination of the invariant sets associated to a discrete
dynamical system is enough information to ascertain the kind of
orbit determined by $x_0$ and $\mu$. There are three types of
invariant sets:
\begin{enumerate}
    \item \emph{Fixed point}. The invariant set $\Gamma$ is a
    fixed point when it contains only one point $x^*$ such that if
    $x_0=x^*$, then $x_k=x^*$ for any $k \in \mathbb{N} \cup \{0\}$.

    \item \emph{Periodic orbit}. In this case, the invariant set
    is a set of $p$ points $\left\{x_1,x_2,\ldots,x_p\right\}$ where
    $x_i=g(\mu,x_{i-1})$ and $x_1=g(\mu,x_p)$.

    \item \emph{Strange attractor}. The main characteristics of
    this kind of invariant set are:
        \begin{itemize}
            \item \emph{Sensitivity with respect to the initial
            conditions}. The orbits associated to two arbitrarily
            close initial conditions are totally different.
            \item \emph{Sensitiviy with respect to the external parameter or set of
            parameters}. Two orbits generated from the same initial
            condition and two arbitrarily close set of values for the external
            parameter(s) are totally different.
            \item \emph{A non-integer dimension}. A strange
            attractor can not be represented neither by a curve nor by a
            surface. For that reason, a new dimension concept is
            introduced. This is named \emph{Haussdorf dimension}.
            \item \emph{Ergodicity}. The orbit associated to a
            certain initial condition included in an small interval
            $I \subset \Gamma$ covers the whole invariant set. Moreover, for other
            small interval $J\subset \Gamma$ there exists a point in $I$ such that
            it determines an orbit that is partially included in the interval $J$.
            Hence, the orbit generated from any initial condition in
            the invariant set is as close to any point
            inside the strange attractor as desired.
        \end{itemize}
\end{enumerate}
Those discrete dynamical systems possessing the last type of
invariant set show a very special behavior when the initial
condition and the parameter(s) are selected to be inside the strange
attractor. This behavior is called chaotic and so these discrete
dynamical systems are named \emph{chaotic discrete dynamical systems
or chaotic maps}. Finally, the chaotic behavior is the reason why
chaotic maps are so attractive for cryptographers. Indeed, the
sensitivity of chaotic maps with respect to the initial conditions
and the control parameter(s) reminds the dependency of cryptosystems
with respect to the secret key. Furthermore, the ergodicity property
of chaotic maps implies that the probability distribution function
associated to the chaotic orbits is independent of the initial
condition and parameter or parameters values, and it also indicates
that two orbits in close proximity in a certain time lead to very
different behaviors after a transient time. In other words, the
ergodic behavior of chaotic maps could be used as the support of
both confusion and diffusion processes required by a secure
cryptosystem.

\begin{figure}[!htbp]
    \centerline{%
    \includegraphics{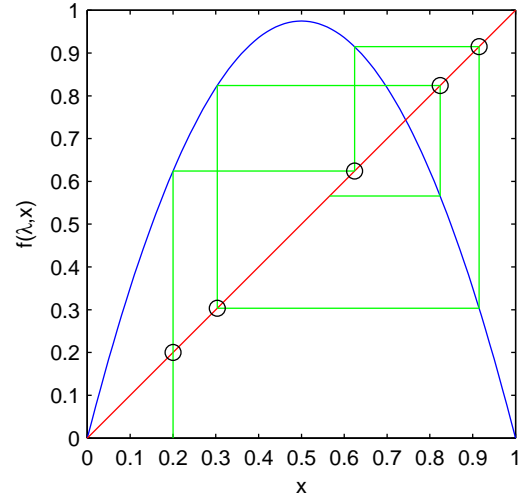}
    }
    \caption{Iteration function associated to the logistic map.}
    \label{figure:iterationFunction}
\end{figure}

\section{The logistic map}
The logistic map is a chaotic map and, having in mind the previous
comments, it can be considered for the design of new digital chaotic
cryptosytems. However, the specific characteristics of the logistic
map make its extensive use not advisable with that purpose. In order
to clarify this assertion, the dynamical characteristics of the
logistic map are going to be thoroughly analyzed through the paper.
First of all, the logistic map is defined by the next equation
\begin{equation}
    x \mapsto f(\lambda,x)= \lambda x (1-x),
    \label{equation:logistic}
\end{equation}
where $f(\lambda,x): U \rightarrow U$, $U=[0,1]$, $\lambda \in
V=[0,4]$ and $x_n \in U$. Figure \ref{figure:iterationFunction}
illustrates the iteration process to generate the orbit of the
logistic when the initial condition $x_0$ is equal to $0.2$ and
$\lambda=3.9$. On the other hand, the function $f(\lambda,x)$
reaches a maximal value for $x=x_c=0.5$ and it is monotonically
increasing for $x<x_c$ and monotonically decreasing for $x
>x_c$. In other words, the logistic map is a \emph{unimodal map}.

\begin{figure}[!htbp]
    \includegraphics{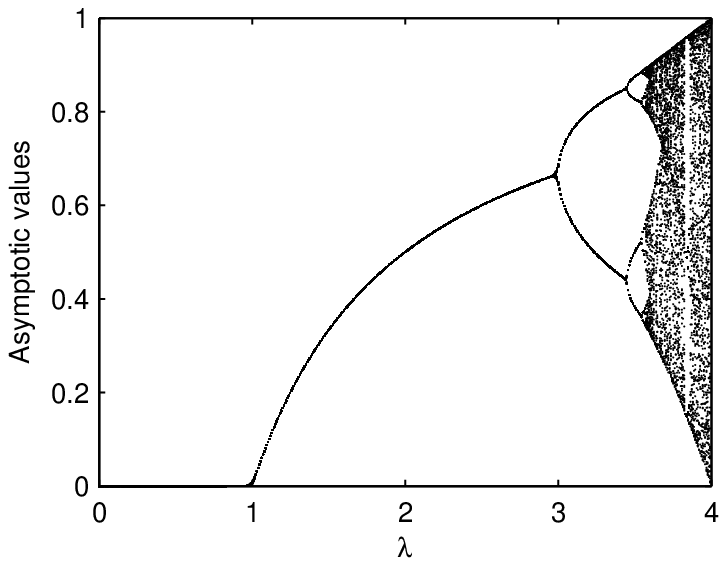}
    \caption{Bifurcation diagram of the logistic map.}
    \label{figure:bifurcationDiagram1}
\end{figure}

The qualitative analysis of the logistic map can be carried out by
the study of its bifurcation diagram. The bifurcation diagram
represents the asymptotic behavior of a discrete dynamical system.
It is calculated iterating the iteration function linked to the
dynamical system for a sufficiently large set of values of the
control parameter(s) and a certain initial condition. In the case of
unimodal maps, the selected initial condition is usually the
critical point. Therefore, to calculate the bifurcation diagram of
the logistic map, Eq.~\eqref{equation:logistic} is iterated from
$x_0=0.5$ for different values of $\lambda$. After a number of
transient iterations, and for each considered value of $\lambda$, a
long enough orbit is recorded and subsequently plotted.
Fig.~\ref{figure:bifurcationDiagram1} has been generated according
to this procedure. A detailed analysis of this bifurcation diagram
leads to the following conclusions about the logistic map:
\begin{itemize}
    \item It has an asymptotic stable fixed point equal to $0$
    for $\lambda \in (0,1)$.
    \item It has an asymptotic stable fixed point equal to
    $\frac{\lambda-1}{\lambda}$ for $\lambda \in (1,3)$.
    \item For $\lambda \in (3,3.57)$ there exist periodic attractors
    of period $2^m$ for $m=1,2,\ldots$. This region is known as the
    doubling-period cascade.
    \item For $\lambda > 3.57$ the behavior of the logistic map is
    chaotic interrupted by small regions where the existing
    attractor is periodic and not strange. These regions are known
    as periodic windows.
\end{itemize}

\subsection{Considerations about the key space determined by the
logistic map} 
\begin{figure}[!htbp]
    \includegraphics{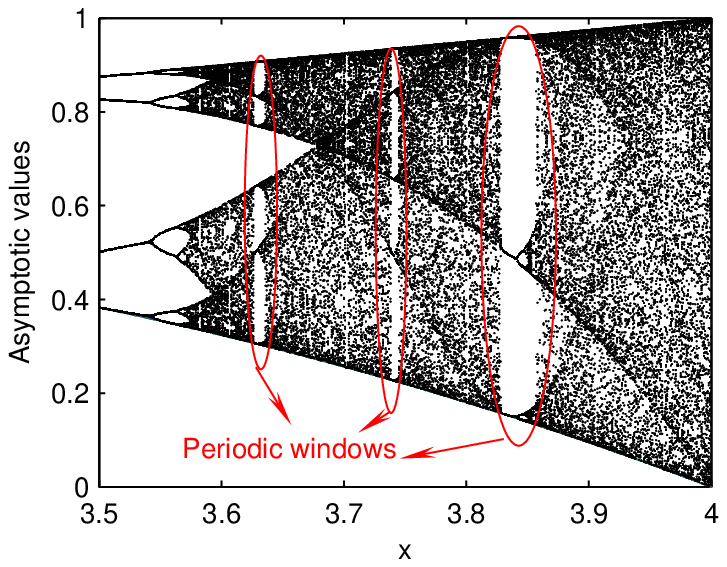}
    \caption{Bifurcation diagram of the logistic map.}
    \label{figure:bifurcationDiagram2}
\end{figure}

In this section it is assumed we are dealing with a
chaotic cryptosystem whose secret key contains the control parameter
$\lambda$ \cite{bap, wong01, wong02,wong03,pareek03,ling07}. As it
was indicated in the previous section, if a chaotic behavior is
demanded for the logistic map, $\lambda$ is required to be greater
than $3.57$. Nonetheless, this is a necessary but not sufficient
condition for the chaoticity of the logistic map, as it was
emphasized above and as Fig.~\ref{figure:bifurcationDiagram2} shows.
This graphic underlines a very important problem when selecting
proper values for $\lambda$ in the case of the logistic map. The
existence of periodic windows must be avoided, since otherwise the
ciphertext would not be random-like and resulting in an inefficient
encryption procedure. This problem has been reported and exploited
for cryptanalysis purposes in \cite{alvarez03a,arroyo07c}.

\begin{figure}[!htbp]
    \includegraphics{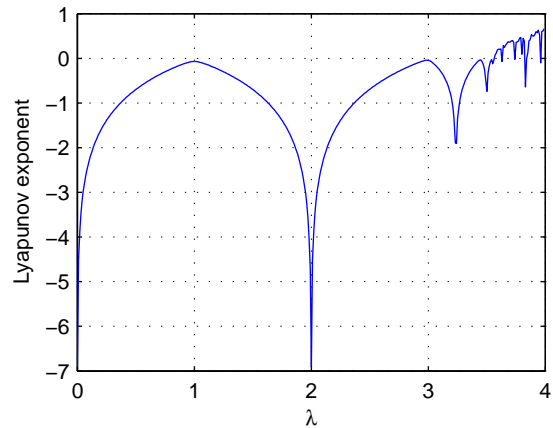}
    \caption{Lyapunov exponent for the logistic map.}
    \label{figure:lyapunovExponent}
\end{figure}

The Lyapunov exponent (LE) \cite{pastor97a} is a very useful tool to
determine if a certain value of $\lambda$ makes the logistic map
evolve chaotically. It is calculated using the following equation
\begin{equation}
    LE(\lambda)=\lim_{N \rightarrow
    \infty}\frac{1}{N}\sum_{k=0}^{N-1}\ln|f(\lambda,x)^\prime(x_k)|.
\end{equation}
Fig.~\ref{figure:lyapunovExponent} depicts the LE for the logistic
map. Those values of $\lambda$ so the LE is greater than zero are
the ones to be selected for cryptographical issues, since they imply
a chaotic behavior. However, the LE does not grab all the periodic
windows, as it was emphasized in \cite{pastor97a}. Therefore, it is
advisable to asses the different values of $\lambda$ using other
entropy measures as the conditional entropy \cite{steuer01}, the
Wavelet-Entropy \cite{Rosso01} or the discrete entropy
\cite{amigo07}. Anyway, the key space related to the control
parameter $\lambda$ undergoes an important stretch due to the
existence of periodic windows, which also makes more difficult the
key selection by the users and thus the cryptosystem implementation.
This last fact should be avoided when designing a new cryptosystem
\cite[Rule 3]{Alvarez06a}, and so the logistic map should be
replaced by another chaotic map that makes easier either the key
selection or the hardware implementation. From this point of view,
Piecewise Linear Maps \cite{li05} should be considered as an
alternative to the logistic map. Indeed, Piecewise Linear Maps have
a Lyapunov exponent always positive which implies that their
bifurcation diagram does not include any periodic window, and they
are the simplest chaotic system from the implementation viewpoint.

\subsection{Considerations about the ergodicity of the logistic map}
\begin{figure*}[!htbp]
    \subfigure[$\lambda=4$]{\includegraphics{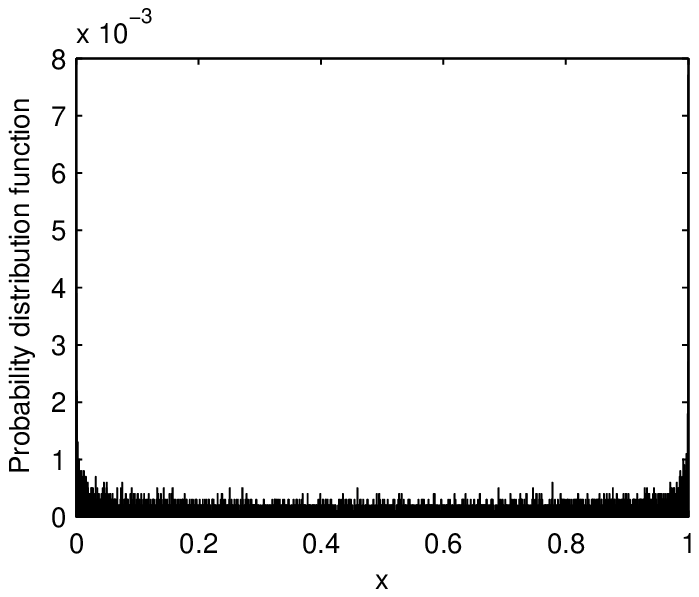}}
    \subfigure[$\lambda=3.8947219769124$]{\includegraphics{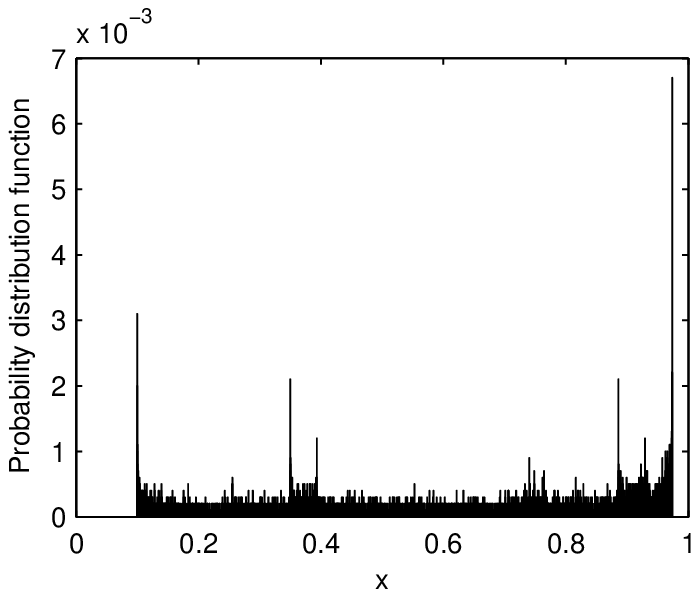}}
    \caption{Probability distribution function of the logistic for different values of the control parameter.}
    \label{figure:pdf}
\end{figure*}

The performance of the Baptista's type chaotic cryptosytems
\cite{bap,wong01,wong02,wong03,huang05,Wei2006,wei06b,oliveira08}
depends on the capacity of the logistic map of generating orbits
with a uniform probability distribution function. Indeed, Baptista's
cryptosystem divides the phase space of the logistic map into a
number of subintervals equal to the cardinality of the alphabet
associated to the messages to be encrypted. In this fashion, each
interval is associated to a character or component of the input
space. Each time a plain character is going to be encrypted, the
logistic map is iterated using the value of $\lambda$ included in
the secret key until the orbit lands on the interval linked to the
plain character. As a result, the efficiency of the encryption
process requires that the probability of landing in a subinterval to
be uniform. This is the reason why the subintervals associated to
the plain alphabet in \cite{bap} are placed in $[0.2,0.8]$ instead
of $[0,1]$. Nevertheless, if the probability density distribution of
the logistic map is analyzed (Fig.~\ref{figure:pdf}), one can
conclude that the most visited places of the phase space of the
logistic map are those closer to the extremes \cite[p.
174]{alvarez03a}, which implies that the encryption speed is very
slow. Again, it is highly advisable to replace the logistic map by
another map with uniform probability distribution function.

\subsection{Considerations about the orbit of the logistic map as ciphertext}
A very important step in the design of a cryptosystem consists in
deciding what is going to be the ciphertext. Indeed, chaotic
cryptosystems transform the input information according to a chaotic
orbit and the resulting ciphertext can not contain enough
information to guess either the plaintext or the secret key. If the
ciphertext contains values of the chaotic orbit it might be possible
to infer the value of the control parameter. This happens with the
cryptosystems described in \cite{pisarchik06} and in \cite{ling07}.
In \cite{ling07} the ciphertext is simply the result of adding the
plaintext to a chaotic orbit. Therefore, it is possible to get the
chaotic orbit by a known-plaintext attack and, after that, estimate
the value of the control parameter using
Eq.~\eqref{equation:logistic} \cite{arroyo07c}. Even if the
ciphertext is obtained by a random sampling of a chaotic orbit as in
\cite{pisarchik06}, it is possible to estimate the value of the
control parameter. As Fig.~\ref{figure:iterationFunction} informs,
the maximum value of the iteration function of the logistic map is
reached for $x_n=0.5$, which implies that the maximum possible value
contained in an orbit is $\lambda/4$. As a result, if one has access
to a sufficiently large sample of ciphertext derived from the
cryptosystem described in \cite{pisarchik06}, then one can estimate
the value of $\lambda$ using the maximum value contained in that
portion of ciphertext. Since this kind of problem is a consequence
of the logistic map having a maximum value dependent on the control
parameter, any chaotic map showing this property is also equally
vulnerable. This is the case of the Mandelbrot map and the tent map,
which are also unimodal maps with either a minimum or maximum value
dependent on the control parameter value. In \cite{alvarez00} this
feature is used to cryptanalyze the cryptosystem described in
\cite{alvarez99}, which it is based on the tent map instead of the
logistic map.

\subsection{Considerations about the symbolic dynamics of the logistic map}
The logistic map is an unimodal map whose critical point is
independent of the control parameter value. A first consequence of
this fact is that it is possible an attack similar to that described
in \cite{alvarez03a}. This attack is carried out on the Baptista's
cryptostem and is based on the theory of symbolic dynamics applied
to unimodal maps \cite{metropolis73}. The symbolic sequence
associated to an orbit of an unimodal map is a binary sequence
obtained by comparing each value of the orbit to the value of the
critical point. Let $x_k$ be a value of the orbit and $x_c$ be the
value of the critical point. The symbol corresponding to $x_k$ is
$b_k$ and it is given by
\begin{equation}
    b_k=\left\{ \begin{array}{cc}0 &\textrm{if} \ \ \ x_k<x_c,  \\
    1 & \textrm{if} \ \ \ x_k\ge x_c,\end{array}\right.
\end{equation}
In \cite{alvarez03a} it is explained how a known-plaintext attack
can be used to reconstruct a symbolic sequence associated to the
secret value of $\lambda$. This symbolic sequence can be used to
estimate either the initial condition or the control parameter that
were employed to generate it \cite{wu04,physcon07}. A way to avoid
this problem is to use another map with a critical point dependent
on the value of the secret key or a non-unimodal chaotic map.

\subsection{Considerations about the return map}
\begin{figure}
    \includegraphics[scale=0.6]{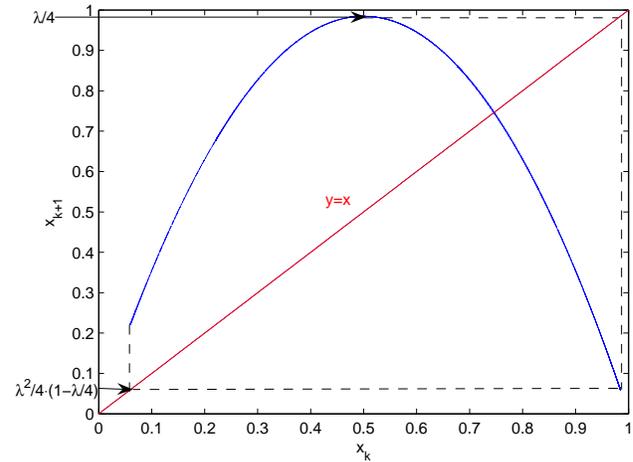}
    \caption{Return map of the logistic for $\lambda=3.9384739$.}
    \label{figure:returMap}
\end{figure}

When observing a dynamical system, a discrete time sequence is
obtained. Let $\left\{x_k\right\}$ be the sequence resulting of the
mentioned observation process. The relationship between two
consecutive values of this discrete time sequence is called
\emph{return map} and it can be used to guess some characteristics
of the underlying dynamical system. Fig. \ref{figure:returMap} shows
the return map of the logistic map, which provides enough
information to get the value of the control parameter. As an
example, in Fig. \ref{figure:returMap} the return map for
$\lambda=3.9384739$ is depicted, showing that for a certain value of
the control parameter the return map is defined in an interval whose
minimum and maximum bounds are $\lambda^2/4\cdot (1-\lambda/4)$ and
$\lambda/4$ respectively. This means that the control parameter can
be obtained through the return map, as for example was done in
\cite{skrobek07}, where a chosen ciphertext attack on the Baptista's
cryptosystem is used to get a discretized return map and
consequently estimate the value of the control parameter. Therefore,
it is advisable to look for a chaotic map with a return map
independent of the value of the control parameter or parameters and
with $x_{k+1}$ being the image of any $x_k$ contained in the phase
space.
\subsection{Considerations about the statistical complexity of the
logistic map} 
\begin{figure}[!htbp]
    \includegraphics{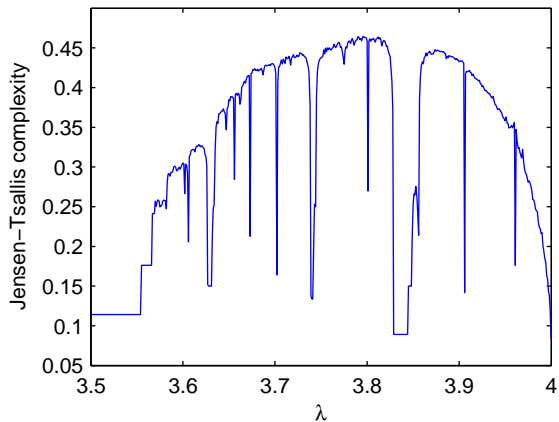}
    \caption{Statistical complexity Jensen-Tsallis of the logistic map for different values of $\lambda$.}
    \label{figure:complexity}
\end{figure}

As detailed in \cite[Table 1]{Alvarez06a}, another
important feature of chaotic maps is their structure complexity.
Chaotic maps are described mathematically in a very simple way but
their behavior is very complex. The complexity of any system can be
measured by using statistical tools \cite{martin06}. In
Fig.~\ref{figure:complexity} appears the Jensen-Tsallis statistical
complexity \cite{martin06} of the logistic map versus the control
parameter. It shows how the statistical complexity is very related
to the value of the control parameter in the chaotic region. In
fact, the complexity is almost a bijective application of the
control parameter. This is something that could be exploited by a
cryptanalyst and thus it is better to use a chaotic map with a
non-bijective relationship between the statistical complexity and
the control parameter.

\section{Conclusions}
In this paper the logistic map has been thoroughly analyzed and many
disadvantages of its application for cryptographical issues has been
emphasized. The logistic map is an unimodal map with an associated
non-uniform probability distribution function, which possesses a
critical point independent of the value of the control parameter and
whose statistical complexity decreases as the control parameter
increases. For all these reasons, the generalized use of the
logistic map for the design of new cryptosystems has to be
discouraged, being Piecewise Linear Maps a good alternative.

\section*{Acknowledgment}
The work described in this paper was supported by \textit{Ministerio
de Educaci\'on y Ciencia of Spain}, research grant SEG2004-02418,
\textit{Ministerio de Ciencia y Tecnolog\'{i}a} of Spain, research
grant TSI2007-62657 and \textit{CDTI, Ministerio de Industria,
Turismo y Comercio of Spain} in collaboration with Telef\'onica I+D,
Project SEGUR@ with reference CENIT-2007 2004.

\bibliographystyle{IEEEtran}

\end{document}